\documentclass[conference, letterpaper]{IEEEtran}

\pdfoutput=1

\ifCLASSINFOpdf
\else
\fi
\usepackage[cmex10]{amsmath}
\ifCLASSINFOpdf
   \usepackage[pdftex]{graphicx}
\else
\fi

\usepackage{verbatim} 
\usepackage{multicol}  
\usepackage{wrapfig}

\usepackage{amsfonts}
\usepackage{float}   

\usepackage{mathtools}
\usepackage{color}
\usepackage{fancyhdr}
\usepackage[caption=false,font=footnotesize]{subfig}

\renewcommand{\thispagestyle}[2]{} 

\fancypagestyle{plain}{
        \fancyhead{}
        \fancyhead[C]{first page center header}
        \fancyfoot{}
        \fancyfoot[C]{first page center footer}
}
\newcommand{\n}{|\!|}

\newenvironment{thm}{{\bf Theorem:  }}

\newcommand{\be}[1]{\begin{equation}\label{#1}}

\newcommand{\ee}{\end{equation}}
\newcommand{\bc}{\begin{center}}
\newcommand{\ec}{\end{center}}
\newcommand{\ba}[1]{\begin{array}{#1}}
\newcommand{\ea}{\end{array}}

\newcommand{\bit}{\begin{itemize}}
\newcommand{\eit}{\end{itemize}}
\newcommand{\ben}{\begin{enumerate}}
\newcommand{\een}{\end{enumerate}}

\author{
\IEEEauthorblockN{Abdolrahman Khoshrou}
\IEEEauthorblockA{
Centrum Wiskunde \& Informatica\\
Science Park 123, 1098 XG \\
Amsterdam, The Netherlands\\
Email: a.khoshrou@cwi.nl}
\and
\IEEEauthorblockN{Eric J. Pauwels}
\IEEEauthorblockA{
Centrum Wiskunde \& Informatica\\
Science Park 123,1098 XG \\
Amsterdam, The Netherlands\\
Email: eric.pauwels@cwi.nl
}
}

\title{{Data-driven pattern identification and outlier detection in time series}}

\begin{document}

%


\maketitle
\begin{abstract}
We address the problem of data-driven pattern identification and outlier detection in time series. 
To this end, we use singular value decomposition (SVD) which is a well-known technique to compute a low-rank approximation for 
an arbitrary matrix. 
By recasting the time series as a matrix it becomes possible to use SVD to highlight the underlying patterns and periodicities.  
This is done without the need for specifying user-defined parameters. 
From a data mining perspective, this opens up new ways of analyzing time series  in a data-driven, bottom-up fashion.  
However, in order to get correct results, it is important to understand how the SVD-spectrum of a time series is influenced by various characteristics of the underlying signal and noise. 
In this paper, we have extended the work in earlier papers by initiating a more systematic analysis of these effects.  We then illustrate our findings on some real-life data. 
\end{abstract}
\IEEEpeerreviewmaketitle
\section{Introduction} \label{sct:intro}
\subsection{Motivation}
Since the gathering of the sensor data has become relatively cheap and straightforward, nowadays it is common to collect detailed information about all sorts of processes and services that  take place in factories, infrastructural networks and public spaces. 
In many of these applications (especially those related to human activities), there is a multitude of  time series in which a pronounced but relatively short periodicity (e.g. daily pattern) is  superimposed on a slower, more global trend. 
If this underlying trend is simple or regular, classic detrending algorithms (e.g.~\cite{wu2007trend,kantelhardt2002multifractal}) can be applied to remove it.  
However, these techniques fall short if it is difficult to identify clear underlying patterns. 
In this paper we propose to use singular value decomposition (SVD) as a way to extract regular periodic patterns in a data-driven fashion. 

The basic idea is fairly straightforward and was first proposed in~\cite{kanjilal1994singular}. 
Let us suppose that one has a (1-dim) periodic time series $\mathbf{x} = (x_1, x_2,\ldots, x_n)$ that has a known period $p$.  
We can then reshape this time series into a matrix using the first $p$ observations (i.e. $x_1$ through $x_p$) to construct the first column, the second set of $p$ observations ($x_{p+1}$ through $x_{2p}$) as the second column, and so on. 
Assuming that the length $n$ of the time series is an integer multiple  (say $q$) of $p$  (i.e. $n = pq$), this reshaping 
results in a $p \times q$ matrix $A$. 
If the time series is perfectly periodic and noiseless, this matrix $A$ has rank 1, since all the columns are linearly dependent.
This means that $A$  can be expressed 
as the product of a single 
($p$-dimensional) column $\mathbf{u}$ and 
($q$-dimensional) row $\mathbf{v}^T$: 
\begin{equation}
     A = \sigma_1 \mathbf{u} \mathbf{v}^T   
     \label{eq:rk_1}
\end{equation}  
where $\sigma_1>0$ is a scaling factor to ensure the 
normalization  $ \n\mathbf{u}\n = \n\mathbf{v}\n = 1$. 
In fact, in the case of identical columns, 
$\mathbf{v} = \mathbf{1}_q $  (i.e. all $v$-entries are equal to 1), 
while $\mathbf{u}$ will be equal to the common column. 

Obviously, the above represents an extreme case where all the 
singular values beyond the first one vanish. 
If we sprinkle a bit of noise onto the time series, the columns 
in $A$ will no longer be identical, but still very similar.  
As a consequence,  
the expression in~\eqref{eq:rk_1} will still hold to a very good approximation. 
This observation is the motivation for the introduction of singular value decomposition 
(SVD) which we will briefly recapitulate below.   
\subsection{SVD recapitulation and some notation}
The basic result that we will use throughout the paper is the following well-known theorem. 

\noindent
\begin{thm}  {\bf Singular Value Decomposition (SVD)} 
Given an arbitrary $p \times q$  matrix 
$A \in \mathbb{R}^{p \times q}$, then there exists 
matrices $U$ and $V$ (both with orthonormal columns), and positive 
numbers $\sigma_1 \geq \sigma_2  \geq \ldots \geq \sigma_r $  (where $r = \min(p,q)$), 
such that: 
\begin{equation}
 A   = \sum\limits_{k = 1}^r \sigma_kU_k V_k^T  = USV^T
    \label{eq:svd}
\end{equation}
with $U_k$ and $V_k$ denoting the $k^{th}$ column of $U$ and $V$, respectively, and $S$ is an $p\times q$ matrix for which the numbers $\sigma_k$ 
(the singular values) are placed on the main diagonal. 
For a proof, see e.g.~\cite{strang1993introduction}.
\end{thm}
\noindent
In the remainder of this paper, we will assume that the singular values are arranged in descending order: 
$  \sigma_1 \geq \sigma_2  \geq \ldots \geq \sigma_n \geq 0. $
For a given matrix $A$ we use the notation $\sigma_i(A)$ or $\lambda_i(A)$ to denote 
the $i$-th (ordered) singular or eigen-value, respectively. If there is no danger of confusion, the explicit reference to the matrix will be suppressed.
Recall that there is a useful relationship between the singular values of a matrix $A \in \mathbb{R}^{p\times q}$ and the eigenvalues  of the related matrices $AA^T$ and $A^T A$:
\begin{equation}
\sigma_i(A) = \sqrt{\lambda_i(AA^T)} = \sqrt{\lambda_i(A^T A)} . 
    \label{eq:eigenvals}
\end{equation}
where $i = 1:\min(p,q)$.
This connection will be used extensively in the analysis below. 
\subsection{Applying SVD to time series}
\begin{figure}
    \centering
    \includegraphics[width = 7.4cm,height=4cm]{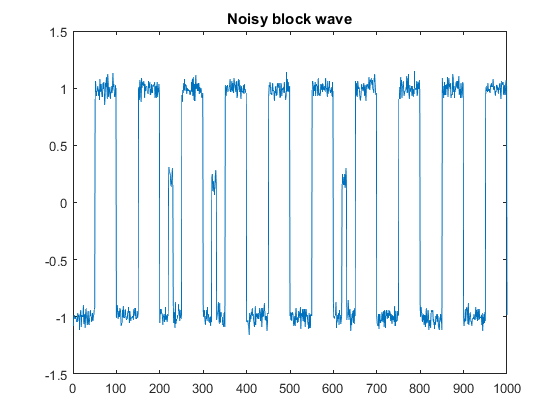}
     \includegraphics[width = 7.4cm,height=4cm]{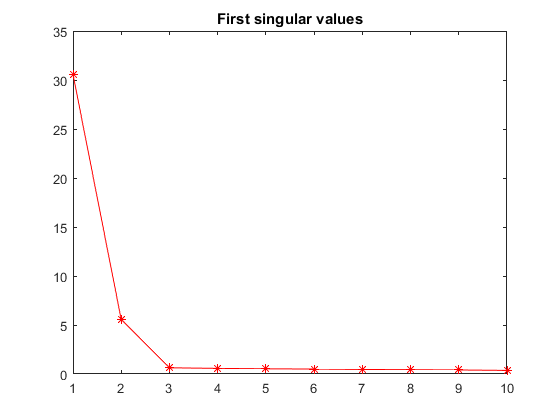}\\
      \includegraphics[width=7.4cm]{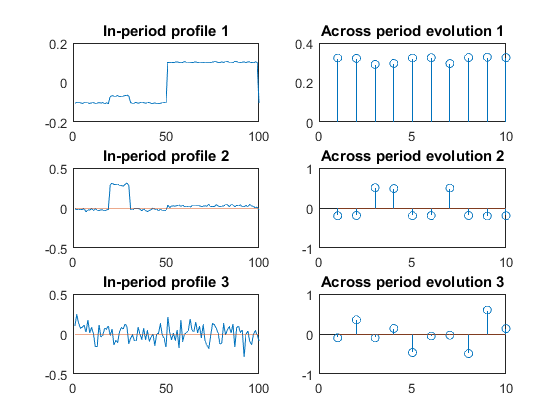}
    \caption{SVD application to pattern-extraction in noisy block signal.  {\bf Top: } Original data 
    of noisy block signal with period 100. 
    In addition to the noise there are three  irregularly occurring 
    spikes. 
    {\bf Middle: } The 10 singular values for SVD with period $p = 100$.  Clearly, only the two first 
    are significant and $\sigma_1 \gg \sigma_2$ confirming that $p=100$ corresponds to a valid periodicity. 
    {\bf Bottom: }  The first three columns of $U$  (left) and $V$ (right). 
    }
    \label{fig:block_signal}
\end{figure}

\begin{figure}
    \centering
    \includegraphics[width=7.4cm]{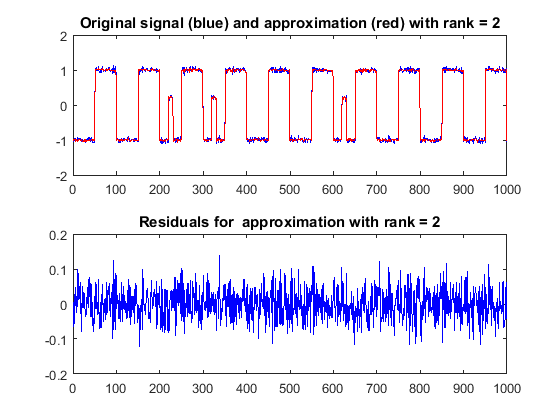}
    \caption{Top: Original (blue) and rank-2 approximation (red) of the block-signal. 
    Bottom:  Residuals with respect to the approximation. }
    \label{fig:block_signal_approx}
\end{figure}

In \cite{kanjilal1995multiple}, the authors draw on SVD to  address the following problems 
for time series:
\begin{enumerate}
    \item {\bf Period extraction:} Given a times series, 
$\mathbf{x}=(x_1,x_2,\ldots x_n)$, reshape it  as a $p \times q$ matrix $A$ (where $p$ ranges 
between some judiciously chosen lower and upper value, 
and $q=floor({ n/p})$, is the nearest integer less than $n/p$. 
The authors then introduce  
the {\it singular value ratio} 
\begin{equation}
    SVR(p) = \sigma_1/\sigma_2
    \label{eq:svr}
\end{equation}
to quantify the dominance of the first singular value over the second. 
High values of the SVR are then considered as an indicator of existence of strong underlying periodicities.    
Plotting $SVR$ as a function of $p$ allows one to spot peaks 
and identify underlying periodicities.
It is not necessarily correct, however, and one must exercise caution when interpreting these graphs, as explained in Section~\ref{sct:mean_shift}. 
\item {\bf Data-driven times series approximation and decomposition:}  
If in the expansion~\eqref{eq:svd} all but the first $r$ singular values are negligible, then 
 truncating the expansion after $r$ terms 
 will still result in an excellent approximation of 
the full matrix $A$ (and corresponding times series).  Furthermore, the columns and rows that 
are retained, can often be interpreted as meaningful patterns (see Fig.~\ref{fig:block_signal}). 
More precisely, for an arbitrary $p\times q$ matrix $A$ (using the notation established above) we know that the ($L_2$) optimal approximation of rank $p < r$ is given by: 
$$ A_p  = \sum_{k=1}^p \sigma_k U_kV_k^T   $$
and the Frobenius norm of the residual is given by 
$$\n A-A_p \n_F^2 = \sigma_{p+1}^2 $$
\end{enumerate}
The gist of these observations is clearly illustrated in Fig.~\ref{fig:block_signal}. 
The top panel shows a noisy block signal of length $n = 1000$ with a pronounced period $p=100$ 
and $q = 10$ full cycles.  
In addition to the noise, there are three  irregularly occurring spikes. 
After rewriting this time series as a $100 \times 10 $ matrix $A$, we apply the SVD algorithm to obtain $A = USV^T$  where $S$ is $100 \times 10$ ``rectangular diagonal'' matrix with the 10 singular values on its main diagonal. 
The middle panel shows those ten singular values, clearly illustrating that all except the first two are negligible, which means that the matrix (and therefore the time series) can be accurately represented  by truncating the expansion in~\eqref{eq:svd} after the first two terms,  
i.e. rank-2 approximation 
(see Fig.~\ref{fig:block_signal_approx}). 

Finally, the bottom panel of Fig.~\ref{fig:block_signal} displays the first three columns of $U$ (left) and $V$ (right), respectively. As they correspond to the most significant singular values, they are most important for the reconstruction of the signal.  
The $U$-columns cover one cycle and can be interpreted as successive profiles needed to reconstruct a generic cycle. In that sense, they are analogous to the various trigonometric basis functions in Fourier analysis.
The $V$-columns, on the other hand, specify the amplitudes with which these basis functions need to be combined in order to reproduce the individual cycles observed in the data. 
Not surprisingly, the main profile ($U_1$ top left) reflects the step-like behaviour seen during each cycle. 
As the amplitude of each of these steps is essentially constant, the 10 $V_1$-entries displayed in the top-right panel show little variation. 
The $U_2$ profile (middle, left) captures the shape of the additional spikes that occur at irregular intervals. 
The positive values in the corresponding $ V_2$-coefficients (middle, right) clearly indicate in which intervals these spikes occur. 
Finally, the erratic appearance of both $U_3$ and $V_3$ are a further indication 
(in line with $\sigma_3 \approx 0$) that all structural information has been extracted from the signal. 

\noindent{\bf Contribution of this paper and overview:}  
The main contribution of this paper is to investigate more systematically the effects of noise (Section~\ref{sct:mean_shift}) and signal levels (Section~\ref{sct:signal_shift}) on the SVD spectrum of a time series with a known period. 
In Section~\ref{sct:cooler}, we show how one can use this decomposition to detect and interpret outliers. 
\section{Impact of signal- and noise-levels on singular values }
As mentioned before,~\eqref{eq:svr} was used to identify underlying periods in~\cite{kanjilal1995multiple}.
However, what was apparently not realized is that the singular values are influenced by relative and absolute levels of noise.  
Failing to recognize this interplay can result in biased or misleading results. 
For this reason, we will review and complement some earlier results. 
\subsection{Singular value spectrum of random matrices }
\begin{figure}
\centering
\includegraphics[width=7.4cm]{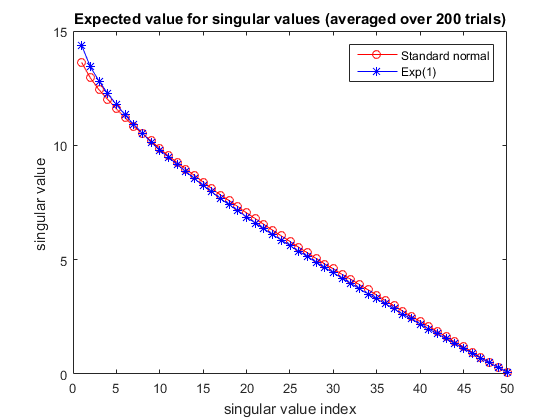}
\caption{Singular values (averaged over 200 trials) for $50\times 50$ 
random matrices generated by drawing i.i.d. entries from the 
standard normal (red) and (shifted to ensure 
zero mean and unit variance) exponential (blue) distributions.  }
\label{fig:singvals_universality}
\end{figure}

In the introductory sections, we assumed that the matrix $A$ was the superposition of some underlying periodic signal and independent noise.  However, to disentangle the impact of signal and noise, we first focus on the effect of pure noise (i.e. random matrices). 
The spectral study of random matrices (i.e. matrices for which the entries are independent, identically distributed (i.i.d.) random variables) has been a very active research domain in recent years and uncovered a number of key insights (see e.g.~\cite{Tao_VanVu_2009,paul2014random,nguyen2014random}).  
One of the more striking results is the emergence of {\it universality} which basically says that as the size of the matrix grows, the distribution of the singular values becomes increasingly  independent of the distribution of the individual entries. 
Put differently, as long as the mean and variance of the noise is kept constant, its actual distribution has very little influence on the distribution of the resulting singular values, assuming the size of the matrix is not too small.  
This surprising result is illustrated in Fig.~\ref{fig:singvals_universality}  where we compare the singular values 
(averaged over 200 trials) of $50\times 50 $ random matrices for two different distributions of the individual matrix entries: 
standard normal and exponential (shifted to become zero-mean). The agreement of the singular values is striking. 

In addition to the above result, we also know that rescaling the variance of 
the entries in 
a zero-mean random matrix induces a  corresponding rescaling of the singular 
values: 
$$  \sigma_i(\alpha A) = \alpha \,\sigma_i(A). $$
This follows immediately from the observation that $\alpha A = U(\alpha S)V^T$.   
In other words, the singular value ratio $SVR = \sigma_1/\sigma_2$ is not affected 
by a uniform increase in the noise variance. 
However, a shift in the mean of the noise does affect the SVR, as will be explained 
in the section below. 
\subsection{Impact of entries mean value} \label{sct:mean_shift} 
In the original papers~\cite{kanjilal1995multiple,ying2004improved}, it was not sufficiently appreciated how a shift in the mean value of the 
time series (the DC component) impacts on the SVR.  
This is important as failure to understand this issue introduces a major bias in the test values and could therefore result in erroneous conclusions.  
To address this issue, we compare the singular values of zero-mean $p\times q$ random matrix $A_0$ and its mean-shifted version: $ A  = A_0 + \alpha$ which is shorthand for $ A = A_0 + \alpha \mathbf{1}_{p\times q} = A_0+\alpha \mathbf{1}_p \mathbf{1}_q^T $. 
Using the connection between singular values and eigenvalues expounded in~\eqref{eq:eigenvals}, we can express any singular value $\sigma(A)$ as: 
$\\
\sigma^2(A) = \lambda(AA^T) \\
= \lambda((A_0 + \alpha \mathbf{1}_p \mathbf{1}_q^T)
(A_0^T+ \alpha \mathbf{1}_q \mathbf{1}_p^T))\\ 
= \lambda \left(A_0A_0^T + 
 \alpha (A_0 \mathbf{1}_q \mathbf{1}_p^T + 
 \mathbf{1}_p \mathbf{1}_q^T A_0^T) + 
 \alpha^2 \mathbf{1}_p \mathbf{1}_q^T \mathbf{1}_q \mathbf{1}_p^T\right) \\
 = \lambda \left(A_0A_0^T + 
 \alpha q(R  \mathbf{1}_p^T + \mathbf{1}_p R^T) + 
 \alpha^2 q \mathbf{1}_p \mathbf{1}_p^T \right)\\
$
where $ R = (1/q)A_0 \mathbf{1}_q $ is a $p\times 1$ column matrix for which each element is the mean of the corresponding $A_0$ row.  
However, recall that the entries of $A_0$ are independent zero-mean  stochastic variables. 
Hence, unless the matrix dimensions are very small, it follows that $R \approx 0$ and can be neglected. 
We therefore derive the approximation: 
\begin{equation}
\sigma^2(A) \approx 
    \lambda \left(A_0A_0^T + 
 \alpha^2 q \mathbf{1}_p \mathbf{1}_p^T \right)
 \label{eq:sigma_approx}
\end{equation}
Next, we make use of the standard results on Rayleigh quotients for eigenvalues which states that the dominant eigenvalue of a symmetric, positive definite matrix $M$ is the solution to the maximization problem: 
$$  \lambda_1 = \max_{\mathbf{x}\neq \mathbf{0}} \left(
\frac{\mathbf{x}^T M  \mathbf{x}}{ \mathbf{x}^T\mathbf{x}} \right)
= \max\limits_{\n\mathbf{u}\n = 1} \, ( \mathbf{u}^T M \mathbf{u}). 
$$
Furthermore, if an unit vector $\mathbf{u}_1$ realizes the above maximum, then the second largest eigenvalue is obtained as the solution of the constrained optimization problem: 
$$
 \lambda_2 
= \max\limits_{\n\mathbf{u}\n = 1}\, ( \mathbf{u}^T M \mathbf{u}) 
\quad\quad s.t. \quad
\mathbf{u} \perp \mathbf{u}_1. 
$$
and so on for the successive eigenvalues. 

\begin{figure}[!h]
    \centering
    \includegraphics[width=8cm]{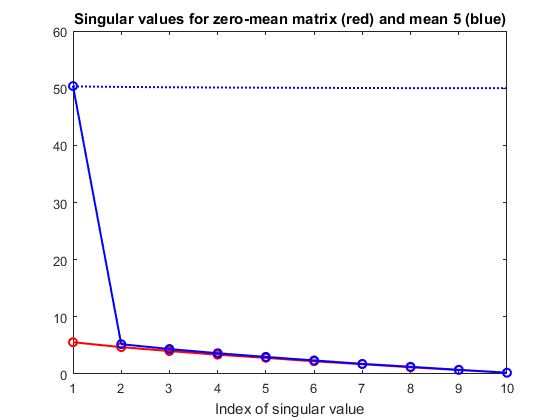}
    \caption{Comparison of the singular values of a matrix ($10 \times 10$) with zero-mean entries (red) and shifted mean ($\alpha = 5$).  The dotted line indicates that (approximate) upper limit based on~\eqref{eq:sigma_1_ineq}. 
    Recall that the entries of the matrix $A_0$ are random numbers, but by shifting the global mean 
    the $SVR = \sigma_1/\sigma_2$ increases, erroneously suggesting that some underlying periodic structure is present.
    }
    \label{fig:singvals_mean_shift}
\end{figure}

Combining this with the approximation derived in~\eqref{eq:sigma_approx}, we get the following approximation 
for the first singular value of $A$:
\begin{align}
\sigma^2(A) & \approx & \max\limits_{\n\mathbf{u}\n = 1}\,
    \mathbf{u}^T\left(A_0A_0^T  + 
 \alpha^2 q \mathbf{1}_p \mathbf{1}_p^T \right) \mathbf{u}\nonumber\\
 &=&  \max\limits_{\n\mathbf{u}\n = 1}\,
    \left( \mathbf{u}^TA_0A_0^T \mathbf{u} + 
 \alpha^2 q \mathbf{u}^T\mathbf{1}_p \mathbf{1}_p^T  \mathbf{u}\right)\nonumber\\
&=&  \max\limits_{\n\mathbf{u}\n = 1} 
\left(\mathbf{u}^TA_0A_0^T \mathbf{u} + \alpha^2q\, (\sum_i u_i)^2\right)
\label{eq:sigma_full}
\end{align}
This derivation shows that 
\begin{equation}
\sigma_1^2(A) 
\leq  \max\limits_{\n\mathbf{u}\n = 1} 
\left(\mathbf{u}^TA_0A_0^T \mathbf{u}\right)  
+\alpha^2 pq  \\
= \sigma_1^2(A_0) + \alpha^2 pq,
\label{eq:sigma_1_ineq}
\end{equation}
since from the Cauchy-Schwartz inequality it follows: 
$$  \left(\sum_i u_i\right)^2 \leq 
\left(\sum_{i=1}^p u_i^2\right) \left(\sum_{i=1}^p 1\right) = p 
\quad \quad \mbox{since}\quad \n u\n = 1.
$$
However, in general the unit vector $\mathbf{u}$ that maximizes the Rayleigh quotient will not necessarily also 
maximize $(\sum u_i)^2 $. In fact, for higher singular values, the number of orthogonal constraints on $u$ increases proportionally, suggesting that on average $\sum u_i \approx 0$, and therefore $\sigma_i^2(A) \approx \sigma_i^2(A_0)$.   
This is indeed exactly what is seen in numerical experiments (Fig.~\ref{fig:singvals_mean_shift}).  
Notice that the first singular value is very close to the maximal value obtained in~\eqref{eq:sigma_1_ineq} 
which is derived if optimizing both terms in~\eqref{eq:sigma_full}, independently and simultaneously was had been done. 

Clearly, failing to remove the mean from a noisy time series would 
inflate the first singular value (and only the first one!) resulting in a upwardly biased value for the singular value ratio (SVR). This would reduce the power of an SVD-method in data mining applications such a blind screening. 
In the next section we will investigate what the impact of genuine underlying periodic signal is. 
\subsection{Impact of the underlying periodic signal} \label{sct:signal_shift}
\begin{figure}
     \centering
     \includegraphics[width=7.4cm]{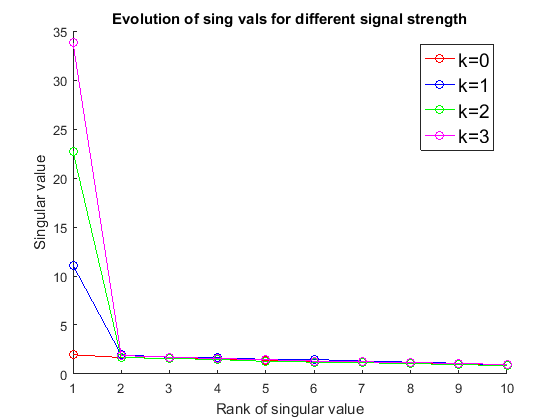}
     \caption{The influence of  the underlying signal strength on the first singular value. The curve for $k=0$ corresponds to  pure noise (no underlying signal).  Notice how increasing the signal strength results in the corresponding increments in the first singular value. 
     }
     \label{fig:signal_approx}
 \end{figure}

Suppose that we have a noisy but perfectly stationary and periodic time series 
$\mathbf{x} = (x_1,x_2, \ldots , x_n)$ with period $p$. For the sake of simplicity, 
we assume that the data cover an integer number $q = n/p $  of periods (cycles). 
As explained in Section~\ref{sct:intro}, we then use the first $p$ observations to create a first column of the matrix $A$, and the observations $x_{p+1},\ldots, x_{2p}$ to create the second column, and so on, until 
we end up with a $p\times q$ matrix $A$. 
If the noise is very small, each column is essentially a copy of the first one and 
we can write: 
$$A \approx \mathbf{a}\mathbf{1}_q^T $$
where the $p\times 1$ column $\mathbf{a}$ represents the data for one period. 
In general, the data is noisy, however, and we model that by adding independent additive noise with variance $\varepsilon^2$: 
$$  A  =   \mathbf{a}\mathbf{1}_q^T  + \varepsilon N. $$
Here $N$ is a $p \times q$  matrix of independent, identically distributed (i.i.d.) 
noise variables with zero mean and unit variance. 
To investigate the behaviour of the singular values we use the fact that 
\begin{eqnarray*}
 \sigma^2(A) &= &\lambda(A^TA)  \\
 &=& \lambda ( (\mathbf{a}\mathbf{1}_q^T  + \varepsilon N )^T (\mathbf{a}\mathbf{1}_q^T  + \varepsilon N)) \\
 &=& \lambda (a^2 \mathbf{1}_q \mathbf{1}_q^T +
 \varepsilon (N^T \mathbf{a} \mathbf{1}_q^T + \mathbf{1}_q \mathbf{a}^T N) +\varepsilon^2 N^TN )
\end{eqnarray*} 
where $a^2 = \mathbf{a}^T \mathbf{a}  = \n \mathbf{a}\n^2$. 
Since the entries of the noise matrix $N$ are independent, zero-mean and unit variance stochastic variables, we can make the following approximation for the $q \times q $  matrix $N^TN $: 
$$  (N^T N)_{ij} = \sum_{k=1}^p N_{ki}N_{kj} \approx  \left\{
\begin{array}{cc}
p    & \mbox{if } i=j  \\
0     & \mbox{if } i \neq j
\end{array}
\right. 
$$
The last approximation is obtained by taking the expected values and using the fact that $E(N_{ki}N_{kj}) = 1 $  if $i=j$, and zero otherwise. 
From this, we conclude that approximately: 
$$   N^TN  \approx pI_q $$
Similarly, because the expectation value of the cross-term vanishes, using the linearity of the expectation operator yields: 
$$ E (N^T \mathbf{a} \mathbf{1}_q^T + \mathbf{1}_q \mathbf{a}^T N) = 
 E (N^T) \mathbf{a} \mathbf{1}_q^T + \mathbf{1}_q \mathbf{a}^T E( N) = 0. 
$$
As a consequence, to a good approximation, the singular values of $A$ can be identified as the eigenvalues of the following matrix: 
$$ \sigma^2(A) \approx \lambda(a^2 \mathbf{1}_q \mathbf{1}_q^T + \varepsilon^2 p I_q) .    $$
The structure of the matrix in the RHS allows us to arrive at some conclusions regarding 
the singular values. Since any vector is an eigenvector of the identity matrix, 
it suffices to focus on the first term which is a rank-1 matrix 
(as the product of a column and a row). 
This implies that all but one eigenvalue vanish, and since $\mathbf{1}_q$ is obviously an eigenvector 
$\left( (\mathbf{1}_q \mathbf{1}_q^T ) \mathbf{1}_q  = 
\mathbf{1}_q (\mathbf{1}_q^T  \mathbf{1}_q) =q  \mathbf{1}_q \right)$
it follows that the maximal eigenvalue (and therefore, singular value) is approximately equal to 
$$ \sigma_1(A) \approx \sqrt{a^2q+\varepsilon^2p} $$
The subsequent singular values correspond to the eigenvectors which are mapped to zero by the 
rank-1 matrix and therefore are not influenced by the $a^2$ term: 
$$  \sigma_i(A) \approx  \sqrt{q}\varepsilon  \quad\quad (\mbox{for }i \geq 2 )  $$
Put differently, these lower ranked singular values  are not influenced by the signal $\mathbf{a}$, 
just by the noise. 
Notice also that the difference between the first and the subsequent singular values grows proportional to $\sqrt{q}$, as it means that the more cycles that are present in the data, the  more pronounced the difference. 
Furthermore,  in many cases the noise-level $\varepsilon^2$ can be neglected with respect to the strength of the signal ($a^2$), resulting in a further approximation: 
$$ \sigma_1(A) \approx \sqrt{q}a. $$
This is illustrated in  Fig.~\ref{fig:signal_approx} where we took a fixed noise-level $\epsilon = 0.2$ and a signal strength $ a$  which is a multiple of some basic level  $a_0 = \sqrt{12.5}$ and $a =k a_0$ with  $k =0, 1, 2, 3$.  
The number of full cycles in each case was equal  to $q=10$. We therefore expect the first singular value for each of these signal levels to be roughly equal to $\sqrt{q}\, a_0 k \approx 11.2k$. 
It is important to realize that this observation  is different from the result in Section~\ref{sct:mean_shift} where the first singular value was affected by a shift in the mean noise level. In this case, the mean $ (1/p) \sum_i a_i$ of the periodic  signal $\mathbf{a}$ can still be zero, but it is its $L_2$ norm ($a^2 = \n\mathbf{a}\n^2 $) that is seen to affect the first singular value. 
\section{Application: Data-driven outlier identification} \label{sct:cooler}
\begin{figure*}
    \centering
    \includegraphics[width=14cm,height=5cm]{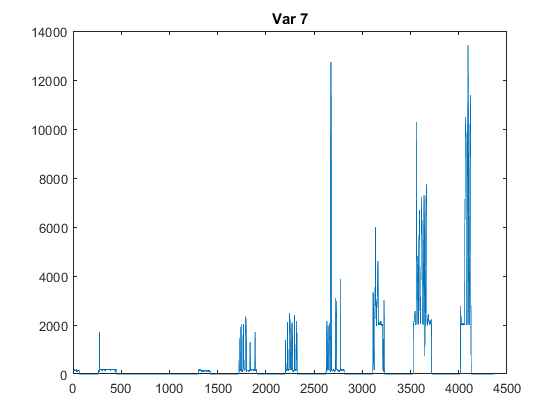}
    \includegraphics[width=14cm,height=7.5cm]{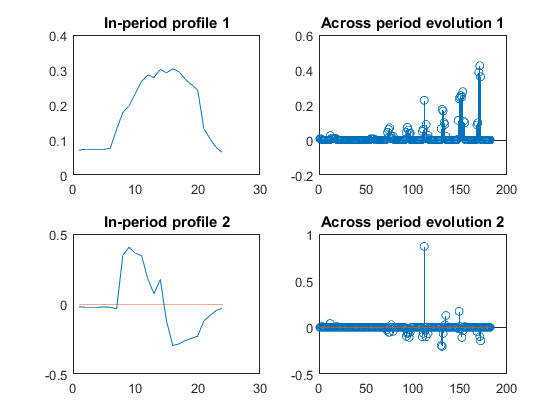}
    \caption{Top: Raw data:  Hourly power consumption of 
    cooler during first 6 
    months of the year. Bursts of activity are interspersed with periods of non-activity. 
    Simple thresholding of the data would suggest that there are a number of 
    unusually high values in the data,  viz. in the 3rd, 5th and (possibly) 6th burst. 
    Bottom two rows: First two U-profiles
    (left) and corresponding V-profiles obtained by SVD.  For more details,  see main text.
    }
    \label{fig:var7_svd_rk2}
\end{figure*}

In the preceding sections we have explored how the singular value spectrum can be used to identify a low-rank approximation of 
a time series and how to avoid misleading biases in the process. 
These low rank approximations provide us with a useful tool to identify and interpret outliers. 
As an illustration, consider the data in the top panel of Fig.~\ref{fig:var7_svd_rk2}  which represents the 
hourly averaged power consumption of an industrial cooler (installed in business offices) over roughly 6 months (January through early July, or 
$n=4368$ data points). 
This cooler works in tandem with two other coolers which explains the burst-like character of the data. 
Since the activity of this cooler is linked to human activity, it 
shows a clear daily periodicity  and we therefore  
performed an SVD with $p = 24$ and $q=n/p=182$.
The plots in the next two rows of Fig.~\ref{fig:var7_svd_rk2}   show 
(left) the first two $U$-columns (24 entries each) 
and (right) the corresponding $V$ -columns 
of length 182 each. 

The two $U$-profiles are plausible: the first captures a (weighted) average of the daily 
activity and therefore shows 
some baseline-activity during the night which then 
ramps up around 8am and  returns to the baseline  
at about 8pm. The additional contribution  encoded in 
the second profile  results in a higher activity 
in the morning, but lower activity in the afternoon.  

\begin{figure*}
    \centering
    \includegraphics[width=14cm]{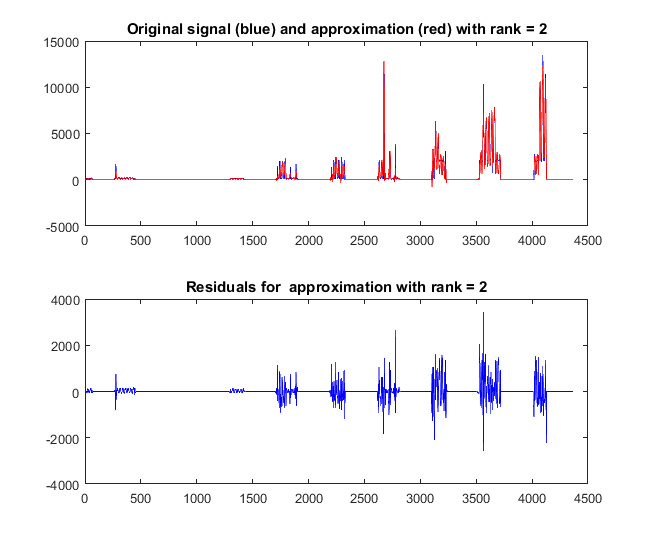}
    \caption{{\bf Top:} Rank-2 approximation(red) of the original signal (blue). 
    {\bf Bottom:}  Corresponding residuals.  Notice that the high peak in the third burst 
    does not yield a high residual because it is adequately  modelled by the extracted profiles. }
    \label{fig:var7_residuals}
\end{figure*}

\begin{figure*}
    \centering
    \includegraphics[width=10cm]{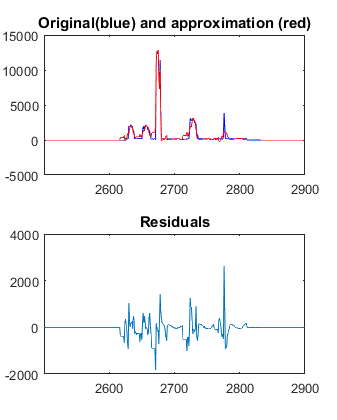}
    \caption{Detail of 3rd "burst" in 
    data of Fig.~\ref{fig:var7_residuals}.  Top: Original (blue) and rank-2 
    approximation (red). Bottom:  Residuals corresponding to top panel.  Notice that the most prominent residual corresponds 
    to a relatively low data peak, which however is unusual 
    in shape.}
    \label{fig:svd_detail_residual}
\end{figure*}

The corresponding $V$-columns on the right specify the appropriate coefficients with which these profiles should be weighted to obtain the approximation (red graph in top panel of Fig.~\ref{fig:var7_residuals}). 
The $V_1$ values roughly mirror the raw data, but the $V_2$ shows a spike that corresponds to the high value in the 3rd burst, indicating that this high value is partly due to an unusually high value in the morning. However, notice that this spike is well modelled by the first two coefficients of the SVD: as a consequence this high value does not result in a corresponding high value for the residual (see bottom panel of Fig.~\ref{fig:var7_residuals} and the zoomed-in version in Fig.~\ref{fig:svd_detail_residual}). 
In fact, the third burst shows a spike in the residuals but this corresponds to a relatively low value, which however is not adequately captured by a combination of the first two $U$-profiles. 

So using this type of analysis we can easily make the distinction 
between high values that are the result of unusual but regular activity 
(encoded in $U$-profiles that correspond to large singular values, 
and possibly lower values that however cannot be 
adequately 
approximated by combining such prominent data-driven profiles (i.e. "real" outliers).  

\section{Conclusion}
In this paper we have argued that the  well-known singular value decomposition (SVD) 
(which is usually applied to matrix problems) can also be successfully  applied to 
identify periodic patterns (profiles) in time series. Furthermore, these profiles 
are completely defined by the data and do not require the specification of 
user-defined parameters, apart from the period (which itself can be estimated 
using this approach). As such, this methodology offers a purely data-driven approach 
to adaptive signal approximation, and based on that, outlier detection. 

Moreover, we have shown that a judicious comparison of the $V$-coefficients and residuals allows one to distinguish between different ways in which 
data-points can be atypical or salient. From a data mining perspective, this opens up new ways of analyzing time 
series  in a data-driven, bottom-up fashion.  
However, it then becomes essential to thoroughly understand how the spectrum of time series is influenced by various characteristics of the signal and noise. 
In this paper, we have extended the work in earlier papers by initiating a more systematic analysis of these effects.  
\section*{Acknowledgment}
The authors would like to acknowledge partial support by the  Dutch TTW-project SES-BE.



%




@article{ying2004improved,
  title={Improved singular value decomposition technique for detecting and extracting periodic impulse component in a vibration signal},
  author={Ying, Liu Hongxing Li Jian Zhao and Liangsheng, Qu},
  journal={Chinese Journal of Mechanical Engineering},
  volume={17},
  number={3},
  pages={1},
  year={2004}
}

@article{kanjilal1995multiple,
  title={On multiple pattern extraction using singular value decomposition},
  author={Kanjilal, Partha Pratim and Palit, Sarbani},
  journal={IEEE transactions on signal processing},
  volume={43},
  number={6},
  pages={1536--1540},
  year={1995},
  publisher={IEEE}
}

@article{howland2004generalizing,
  title={Generalizing discriminant analysis using the generalized singular value decomposition},
  author={Howland, Peg and Park, Haesun},
  journal={IEEE transactions on pattern analysis and machine intelligence},
  volume={26},
  number={8},
  pages={995--1006},
  year={2004},
  publisher={IEEE}
}

@article{choi1989improved,
  title={Improved time-frequency representation of multicomponent signals using exponential kernels},
  author={Choi, H-I and Williams, William J},
  journal={IEEE Transactions on Acoustics, Speech, and Signal Processing},
  volume={37},
  number={6},
  pages={862--871},
  year={1989},
  publisher={IEEE}
}

@book{strang1993introduction,
  title={Introduction to linear algebra},
  author={Strang, Gilbert },
  volume={},
  year={1993},
  publisher={Wellesley-Cambridge Press Wellesley, MA}
}


@article{wu2007trend,
  title={On the trend, detrending, and variability of nonlinear and nonstationary time series},
  author={Wu, Zhaohua and Huang, Norden E and Long, Steven R and Peng, Chung-Kang},
  journal={Proceedings of the National Academy of Sciences},
  volume={104},
  number={38},
  pages={14889--14894},
  year={2007},
  publisher={National Acad Sciences}
}


@article{kantelhardt2002multifractal,
  title={Multifractal detrended fluctuation analysis of nonstationary time series},
  author={Kantelhardt, Jan W and Zschiegner, Stephan A and Koscielny-Bunde, Eva and Havlin, Shlomo and Bunde, Armin and Stanley, H Eugene},
  journal={Physica A: Statistical Mechanics and its Applications},
  volume={316},
  number={1},
  pages={87--114},
  year={2002},
  publisher={Elsevier}
}


@article{Tao_VanVu_2009,
  title={Random Matrices: The distribution of the smallest singular values},
  author={Tao, Terence and Vu, Van H. },
  journal={ArXiv: 0903:0614},
  year={2009},
  publisher={Cornell University Library}
}

@article{kanjilal1994singular,
  title={The singular value decomposition—Applied in the modelling and prediction of quasiperiodic processes},
  author={Kanjilal, Partha Pratim and Palit, Sarbani},
  journal={Signal processing},
  volume={35},
  number={3},
  pages={257--267},
  year={1994},
  publisher={Elsevier}
}

@article{paul2014random,
  title={Random matrix theory in statistics: A review},
  author={Paul, Debashis and Aue, Alexander},
  journal={Journal of Statistical Planning and Inference},
  volume={150},
  pages={1--29},
  year={2014},
  publisher={Elsevier}
}

@article{nguyen2014random,
  title={Random matrices: Law of the determinant},
  author={Nguyen, Hoi H and Vu, Van and others},
  journal={The Annals of Probability},
  volume={42},
  number={1},
  pages={146--167},
  year={2014},
  publisher={Institute of Mathematical Statistics}
}

\begin{thebibliography}{1}
\providecommand{\url}[1]{#1}
\csname url@samestyle\endcsname
\providecommand{\newblock}{\relax}
\providecommand{\bibinfo}[2]{#2}
\providecommand{\BIBentrySTDinterwordspacing}{\spaceskip=0pt\relax}
\providecommand{\BIBentryALTinterwordstretchfactor}{4}
\providecommand{\BIBentryALTinterwordspacing}{\spaceskip=\fontdimen2\font plus
\BIBentryALTinterwordstretchfactor\fontdimen3\font minus
  \fontdimen4\font\relax}
\providecommand{\BIBforeignlanguage}[2]{{%
\expandafter\ifx\csname l@#1\endcsname\relax
\typeout{** WARNING: IEEEtran.bst: No hyphenation pattern has been}%
\typeout{** loaded for the language `#1'. Using the pattern for}%
\typeout{** the default language instead.}%
\else
\language=\csname l@#1\endcsname
\fi
#2}}
\providecommand{\BIBdecl}{\relax}
\BIBdecl

\bibitem{wu2007trend}
Z.~Wu, N.~E. Huang, S.~R. Long, and C.-K. Peng, ``On the trend, detrending, and
  variability of nonlinear and nonstationary time series,'' \emph{Proceedings
  of the National Academy of Sciences}, vol. 104, no.~38, pp. 14\,889--14\,894,
  2007.

\bibitem{kantelhardt2002multifractal}
J.~W. Kantelhardt, S.~A. Zschiegner, E.~Koscielny-Bunde, S.~Havlin, A.~Bunde,
  and H.~E. Stanley, ``Multifractal detrended fluctuation analysis of
  nonstationary time series,'' \emph{Physica A: Statistical Mechanics and its
  Applications}, vol. 316, no.~1, pp. 87--114, 2002.

\bibitem{kanjilal1994singular}
P.~P. Kanjilal and S.~Palit, ``The singular value decomposition—applied in
  the modelling and prediction of quasiperiodic processes,'' \emph{Signal
  processing}, vol.~35, no.~3, pp. 257--267, 1994.

\bibitem{strang1993introduction}
G.~Strang, \emph{Introduction to linear algebra}.\hskip 1em plus 0.5em minus
  0.4em\relax Wellesley-Cambridge Press Wellesley, MA, 1993.

\bibitem{kanjilal1995multiple}
P.~P. Kanjilal and S.~Palit, ``On multiple pattern extraction using singular
  value decomposition,'' \emph{IEEE transactions on signal processing},
  vol.~43, no.~6, pp. 1536--1540, 1995.

\bibitem{Tao_VanVu_2009}
T.~Tao and V.~H. Vu, ``Random matrices: The distribution of the smallest
  singular values,'' \emph{ArXiv: 0903:0614}, 2009.

\bibitem{paul2014random}
D.~Paul and A.~Aue, ``Random matrix theory in statistics: A review,''
  \emph{Journal of Statistical Planning and Inference}, vol. 150, pp. 1--29,
  2014.

\bibitem{nguyen2014random}
H.~H. Nguyen, V.~Vu \emph{et~al.}, ``Random matrices: Law of the determinant,''
  \emph{The Annals of Probability}, vol.~42, no.~1, pp. 146--167, 2014.

\bibitem{ying2004improved}
L.~H. L. J.~Z. Ying and Q.~Liangsheng, ``Improved singular value decomposition
  technique for detecting and extracting periodic impulse component in a
  vibration signal,'' \emph{Chinese Journal of Mechanical Engineering},
  vol.~17, no.~3, p.~1, 2004.

\end{thebibliography}


\newpage
\onecolumn 


\end{document}